\documentclass[a4paper,11pt,notitlepage]{article}
\pdfoutput=1
\usepackage{graphicx}
\usepackage[margin=30mm]{geometry}
\usepackage{setspace}
\usepackage{amsmath}
\usepackage[font={small}, labelfont={bf,small}]{caption}
\usepackage{siunitx}
\usepackage{cite}

\usepackage{xspace}
\usepackage{amssymb}
\usepackage{amsmath}
\usepackage{bm}
\usepackage{siunitx}
\usepackage[acronym,nonumberlist,toc]{glossaries}
\usepackage{xparse}

\usepackage{setspace} 
\DeclareCaptionLabelSeparator{bar}{\,$\boldsymbol{\vert}$\,}
\newcommand{\be}{\begin{equation}}
\newcommand{\ee}{\end{equation}}

\newcommand{\pnt}{\mbox{.}}
\newcommand{\virg}{\mbox{,}}
\newcommand{\ndt}{non-diffusive transport\xspace}
\newcommand{\Ndt}{Non-diffusive transport\xspace}

\DeclareMathOperator{\sign}{sign}
\DeclareMathOperator{\Gamfnct}{\Gamma}
\DeclareMathOperator{\diff}{d\hspace{-0.7pt}}
\DeclareMathOperator{\fracdiff}{D}
\DeclareRobustCommand{\uppartial}{\text{\rotatebox[origin=t]{13}{\scalebox{0.95}[1]{$\partial$}}}\hspace{-0.5pt}}

\newglossarystyle{modsuper}{%
  \glossarystyle{super}%
  
}

\newglossarystyle{mystyle}{%
  \glossarystyle{list}%
}

\makeglossaries

\newacronym{pdf}{PDF}{probability distribution function}
\newacronym{ctrw}{CTRW}{continuous time random walk}
\newacronym{flm}{fLm}{fractional L\'evy motion}

\newacronym{aflm}{afLm}{asymmetric fractional L\'evy motion}

\newacronym{clt}{CLT}{central limit theorem}
\newacronym{lhs}{lhs}{left hand side}
\newacronym{rhs}{rhs}{right hand side}
\newacronym{fbm}{fBm}{fractional Brownian motion}
\newacronym{obm}{oBm}{ordinary Brownian motion}
\newacronym{fgn}{fGn}{fractional Gaussian noise}
\newacronym{gle}{GLE}{generalized Langevin equation}
\newacronym{slm}{sLm}{stable L\'evy motion}

\newglossaryentry{alpha}
{
  name={\ensuremath{\alpha}},
  description={Spatial transport exponent of the fractional L\'evy motion},
  sort=gammaralpha
}
\newcommand{\alflm}{\gls{alpha}\xspace}

\newglossaryentry{beta}
{
  name={\ensuremath{\beta}},
  description={Temporal transport exponent of the fractional L\'evy motion},
  sort=gammarbeta
}
\newcommand{\beflm}{\gls{beta}\xspace}

\newglossaryentry{hurst}
{
  name={\ensuremath{H}},
  description={Self-similarity exponent, also called Hurst exponent},
  sort=hurst
}
\newcommand{\hurst}{\gls{hurst}\xspace}

\newglossaryentry{gammaR}
{
  name={\ensuremath{\gamma_R}},
  description={Radial transport exponent},
  sort=gammar
}

\begin{document}

 \begin{center}
 \LARGE{\textbf{An Introduction to Non-diffusive Transport Models}}
 
\vspace{1cm}

\large
Alexandre Bovet

\vspace{0.2cm}
\normalsize
\textit{Centre de Recherches en Physique des Plasmas,
\'{E}cole Polytechnique F\'{e}d\'{e}rale de Lausanne (EPFL),
1015 Lausanne, Switzerland\\[0.2cm]
Institute for Integrative Biology, ETH Z\"urich, 8092 Z\"urich, Switzerland.}

\vspace{1cm}
\end{center}

\begin{abstract}

The process of diffusion is the most elementary stochastic transport process.
Brownian motion, the representative model of diffusion, played a important role in the advancement of scientific 
fields such as physics, chemistry, biology and finance.
However, in recent decades, non-diffusive transport processes with non-Brownian statistics were observed 
experimentally in a multitude of scientific fields.
Examples include human travel\cite{Brockmann2006,Song2010}, in-cell dynamics\cite{Caspi2000}, the motion of bright 
points on the solar surface\cite{Lawrence2001}, 
the transport of charge carriers in amorphous semiconductors\cite{Scher1975}, the propagation of
contaminants in groundwater\cite{Kirchner2000}, the search patterns of foraging animals\cite{Ayala-Orozco2004} and the 
transport of energetic particles in turbulent 
plasmas\cite{Perri2009,Bovet2014nf,Bovet2014prl,effenberger2014parameter,Bovet2015,Bovet2015phd}.
These examples showed that the assumptions of the classical diffusion 
paradigm\cite{Einstein1905,Smoluchowski1906,Langevin1908}, assuming an underlying uncorrelated (Markovian), Gaussian 
stochastic process, need to be relaxed to describe transport processes exhibiting a non-local character and exhibiting 
long-range correlations.

This article does not aim at presenting a complete review of non-diffusive transport, but rather
an introduction for readers not familiar with the topic.
For more in depth reviews, we recommend the references\cite{Samorodnitsky1994,Metzler2000,Metzler2004,Mainardi2001}.
First, we recall the basics of the classical diffusion model and then we present two approaches of possible
generalizations of this model: the \gls{ctrw} and the \gls{flm}.

\end{abstract}

\tableofcontents

\setlength{\parskip}{1em}

\section{Classical diffusion and the random walk model}
\label{sec:classical_diff}

\begin{figure}[!tb]
\centering
\includegraphics[width=0.6\linewidth]{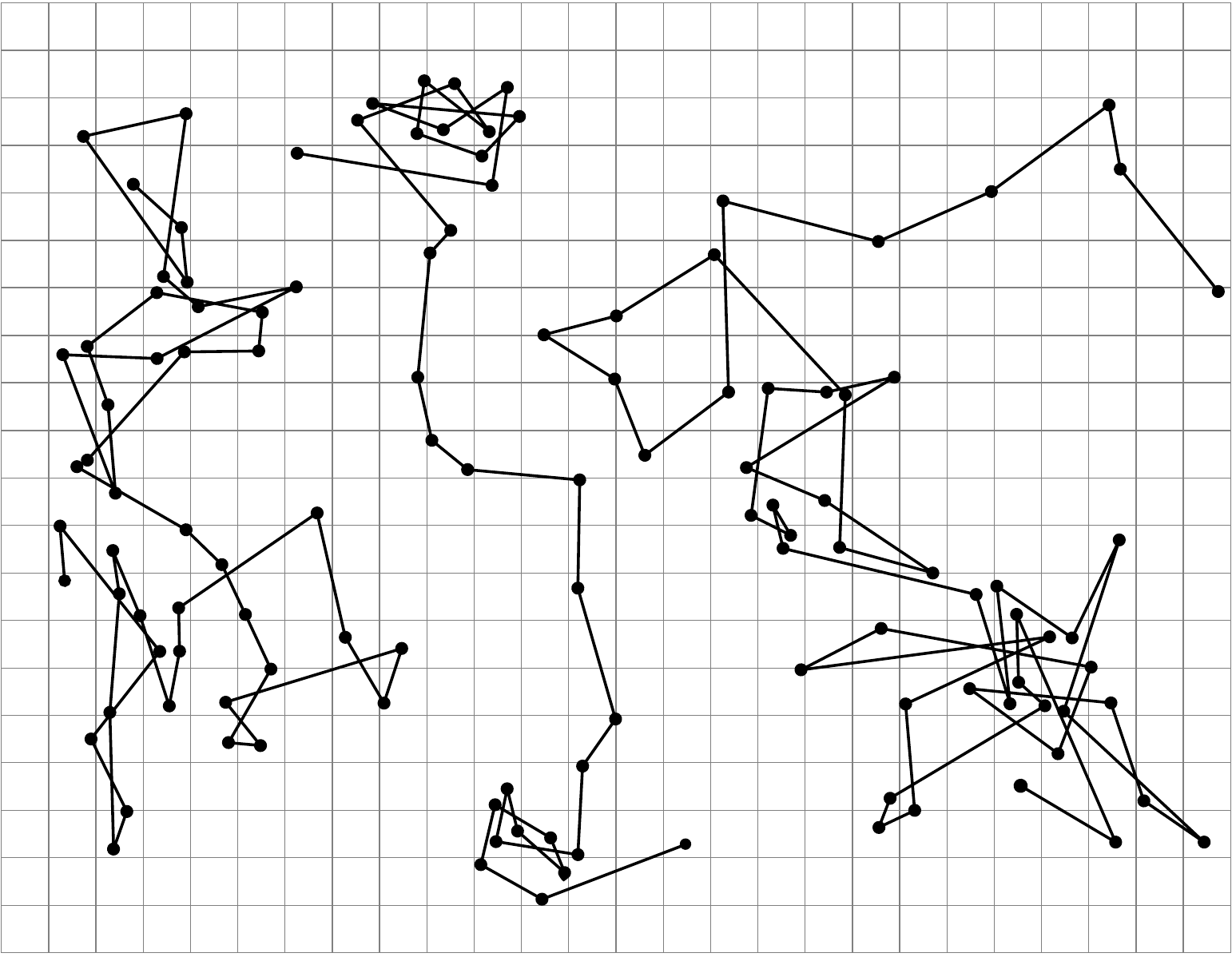}
\caption{Example of trajectories of small particles (radius of \SI{0.53}{\micro m}) in suspension in a fluid measured
by Jean Perrin\cite{Mandelbrot1983fractal}.
The successive positions are marked every \SI{30}{s} and joined by straight lines to guide the eye.
The grid size is \SI{3.2}{\micro m}.} 
\label{fig:PerrinPlot}
\end{figure}

The model of the random walk was first developed by Einstein\cite{Einstein1905} in 1905 and, independently,
by Smoluchowski\cite{Smoluchowski1906} in 1906, to explain the observation made by a Scottish botanist, Robert Brown
(1773-1858), of the random motion of pollen particles in suspension on water. 
The term \textit{Brownian motion} has been coined in honors of Brown to designate the random walk.
In 1913, in his book \textit{Les Atomes}\cite{Perrin1913}, Perrin verified the results of Einstein and Smoluchowski by
measuring with a microscope the displacement of small particles in suspension in a liquid.
Figure \ref{fig:PerrinPlot} reproduces some of his observations.
Using Einstein's theory, he was able to measure the Avogadro Number\cite{Philibert2005}. 
This remarkable success bore the definite proof of the existence of the atom which awarded Perrin the Nobel Prize for
physics in 1926.

Einstein's model of the random walk assumes that each individual particle motion is independent of the other's and that
the displacements of the same particle at different times are also independent, provided that the interval separating 
the different times is not too small.
Next, he introduces a time interval, $\tau$, very small compared to the observation time but sufficiently large so
that the motions between two consecutive time intervals $\tau$ can be considered as independent.
In a one-dimensional (1D) model, considering $n$ particles, during the time interval $\tau$, each particle position 
along the $x$-axis will increase by a value $\Delta$, different for each particle.
The \gls{pdf} of the step sizes $\varphi(\Delta)$ is defined by the following relation: the number $\diff n$ of particle
experiencing a displacement lying between $\Delta$ and $\Delta+\diff\Delta$ is given by
\be\label{eq:stepsizepdf}
\diff n= n\varphi(\Delta)\diff\Delta\pnt
\ee
The step size probability satisfies the relation $\varphi(\Delta)=\varphi(-\Delta)$ and differs from zero only for very
small values of $\Delta$.

Let $f(x,t)$ be the distribution of particles.
Using eq. \ref{eq:stepsizepdf}, the number of particles at time $t+\tau$ found between $x$ and $x+\diff x$ is written
\be
f(x,t+\tau)\diff x=\diff x\int_{-\infty}^{+\infty} f(x+\Delta,t)\varphi(\Delta)\diff\Delta\pnt
\ee
Since $\tau$ is very small, we can write 
\be
f(x,t+\tau)=f(x,t)+\tau\frac{\uppartial f}{\uppartial t},
\ee
and by expanding $f(x+\Delta,t)$ in powers of $\Delta$, we find
\be
f(x+\Delta,t)=f(x,t)+\Delta\frac{\uppartial f(x,t)}{\uppartial x}+\frac{\Delta^2}{2}\frac{\uppartial^2
f(x,t)}{\uppartial^2 x}
+ O(\Delta^3)\pnt
\ee
We note that, after the assumption of independent time steps and symmetrically distributed step sizes, we assume
here that $\tau$ and $\Delta$ cannot take large values in order to perform the two previous expansions. 
This is, as we will see below, a fundamental restriction of the diffusive model which gives it its local character in
time and space.

Since only very small values contribute to it, the expansion can be performed under the integral.
We find
\be
f(x,t)+\tau\frac{\uppartial f}{\uppartial t}=f\int_{-\infty}^{+\infty}\varphi(\Delta)\diff\Delta+
\frac{\uppartial f}{\uppartial x}\int_{-\infty}^{+\infty}\Delta\varphi(\Delta)\diff\Delta+
\frac{\uppartial^2 f}{\uppartial^2 x}\int_{-\infty}^{+\infty}\frac{\Delta^2}{2}\varphi(\Delta)\diff\Delta+
O(\Delta^3)\pnt
\ee
All the terms with odd powers of $\Delta$ vanish due to the fact that $\varphi(\Delta)=\varphi(-\Delta)$.
Taking into account the fact that $\int_{-\infty}^{+\infty}\varphi(\Delta)\diff\Delta=1$, defining the variance of
the step sizes
\be
\left< \Delta^2 \right>=\int_{-\infty}^{+\infty}\Delta^2\varphi(\Delta)\diff\Delta
\ee
and retaining the terms up to $O(\Delta^3)$, we find the well-known diffusion equation
\be\label{eq:diffusion}
\frac{\uppartial f}{\uppartial t}=D\frac{\uppartial^2 f}{\uppartial^2 x}\virg
\ee
where $D=\frac{\left< \Delta^2 \right>}{2\tau}$ is the diffusion coefficient.
The equation of diffusion had already been discovered experimentally by Fick in 1855, but Einstein was the first to
derive it from this physical and atomistic model.

Let us now find the fundamental solution, $G(x,t)$, of eq. \ref{eq:diffusion}, i.e. the solution of the equation 
with
initial condition $G(x,t=0)=\delta(x)$.
This solution is also called the \textit{Green function} or the \textit{propagator} of the equation.
Its convolution with an arbitrary initial condition $f_0(x)$ provides the evolution of the initial condition at all
times $t>0$
\be
f(x,t)=\left(G\ast f_0\right) (x,t)=\int_{-\infty}^{+\infty} G(x-x',t)f_0(x')\diff x'\pnt
\ee
As we will see later, it is interesting to take the Fourier transform in space of eq. \ref{eq:diffusion}
\be\label{eq:ft_diff}
\frac{\uppartial \hat{G}(k,t)}{\uppartial t}=-k^2D\hat{G}(k,t)\virg
\ee
where
\be
\hat{G}(k,t)=\int_{-\infty}^{+\infty}e^{-ikx}G(x,t)\diff x\pnt
\ee
The solution of eq. \ref{eq:ft_diff} is 
\be\label{eq:ft_of_gauss}
\hat{G}(k,t)=Ce^{-k^2Dt}\pnt
\ee
As $\hat{G}(k,0)=\int_{-\infty}^{+\infty}e^{-ikx}\delta(x)\diff x=1$, we find that $C=1$.

We find the solution by taking the inverse Fourier transform of the previous expression
\be
G(x,t)=\frac{1}{2\pi}\int_{-\infty}^{+\infty}e^{-k^2Dt+ikx}\diff k=\frac{1}{2\pi}\int_{-\infty}^{+\infty}e^{-\left(
Dt\left(k - \frac{ix}{2Dt}\right)^2+ \frac{x^2}{4Dt}\right)}\diff k\pnt
\ee
After a little effort, one finds that the solution is the Gaussian, or \textit{normal}, distribution
\be\label{eq:ordinaryBorwnian_propagator}
G(x,t)=\frac{1}{\sqrt{4\pi Dt}}e^{-\frac{x^2}{4Dt}}\virg
\ee
with a variance, or mean-squared displacement, given by 
\be\label{eq:variance_diff_lin}
\left<x^2\right>=2Dt.
\ee
The linear time dependence of the mean-square displacement of diffusive processes is a fundamental results of the
random walk model.

The fact that the distribution of positions of the random walkers is a Gaussian distribution arises naturally from the
\gls{clt}.
Indeed, the position of each particle is a sum of independent and identically distributed steps having the same mean
and the same variance.
Each step being independent, the diffusive process is also memory-less, i.e. Markovian.

\section{\Ndt}
\label{sec:non_diff_transp}

Since its discovery, the model of diffusion and Brownian motion plays a crucial role not only in physics but also 
in biology, chemistry, sociology, economics and finance.
However, numerous examples of transport show a deviation from the diffusive paradigm given by 
eq. (\ref{eq:variance_diff_lin}).
\Ndt is in fact ubiquitous in nature.
Examples ranges from the dispersal of bank notes\cite{Brockmann2006}, the motion of particles inside living 
cells\cite{Caspi2000} or the foraging movements of spider monkeys\cite{Ayala-Orozco2004}.
In plasma physics, examples are also multiple.
The acceleration of electrons and ions by interplanetary shocks in the solar wind\cite{Perri2009}, ion transport across 
the magnetopause\cite{Greco2003}, the motion of magnetic bright points on the solar surface\cite{Lawrence2001} or the 
transport of tracer particles in 3D pressure-gradient driven turbulence\cite{Carreras2001} are evidences 
indicating the presence of \ndt in magnetized plasmas.

\Ndt is characterized by a mean-squared displacement (variance of displacement) of an ensemble of
individuals that does not necessarily scale linearly with time
\be\label{eq:msd_gamma}
\left\langle \left(\bm{r}(t) -\left<\bm{r}(t)\right>\right)^2\right\rangle\propto t^\gamma,
\ee
where $\boldsymbol{r}(t)$ represents the positions of individuals and $\gamma$ is called the \textit{transport
exponent}.
When $\gamma > 1$ or $\gamma < 1$, the transport is called superdiffusive or subdiffusive, 
respectively.
For the special case of classical diffusion, $\gamma = 1$ in accordance with eq. (\ref{eq:variance_diff_lin}).
When $\gamma=2$, the transport is ballistic.
\Ndt is at the heart of many complex systems, such as turbulence, where well-defined scale-lengths 
or time-scales do not exist, and thus transport cannot be modeled as a classical diffusive process.
In these systems, the transport is characterized by the presence of long-term memory and/or non-Gaussian (heavy-tailed) 
\glspl{pdf}.
Mandelbrot coined the terms Noah effect and Joseph effect\cite{Mandelbrot1968noahjosef}, as a reference to the natural 
events experienced by these 
biblical figures, to describe those two effects in the context of hydrology. 
The great flood experienced by Noah and the seven years of abundance followed by seven years of 
famine experienced by Joseph are well known examples that reflects that extreme events with low probability and cycles 
or trend do, in fact, occur in nature.

In order to account for these effects, the hypotheses of the \gls{clt} need to be loosen. 
By removing the restriction on the finiteness of the variance of the random variables (here, the step sizes), we allow 
large fluctuations in the random walk.
The limiting distributions in this case are given by the \textit{generalized central limit theorem}, due to the 
work of L\'evy, Khintchine, Gnedenko and Kolmogorov\cite{Khintchine1936,Bouchaud1990,Samorodnitsky1994} in the 1930.
They are called $\alpha$-\textit{stable distributions} and are presented in appendix \ref{sec:stable_dist}.
They are characterized by their \textit{index of stability}, $\alpha\in(0,2]$, and include the Gaussian distribution as
a special case, for $\alpha=2$.
When $\alpha<2$, they have algebraically decaying heavy tails with exponent $-(1+\alpha)$, and infinite variance.
For this reason, they are particularly interesting to model stochastic processes with high variability,
such as solar flare intermittency\cite{Scafetta2003}.
By removing the hypothesis on the independence of the steps of the \gls{clt}, we allow for long-time correlations in
the Brownian motion.

In the next section, we introduce the model of the \acrfull{ctrw} which is a generalization of the random walk.
We give two notable examples of this model, the \textit{L\'evy-flight} and the \textit{L\'evy-walk}.
Finally, we introduce an other model of \ndt which is based on the Langevin equation of motion and includes long-range
temporal correlations, the \acrfull{fbm} and its generalization to a non-Gaussian, heavy-tailed process, the
\acrfull{flm}.

\subsection{Continuous time random walk}

The model of the \gls{ctrw} was first developed in 1965 by Montroll and Weiss to describe the mobility of charges in
amorphous semiconductors\cite{Montroll1965}.
It has since then found a wide range of applications in physics, chemistry, biology, etc.
The \gls{ctrw} supposes that a particle, also called random walker, makes successive jumps interrupted by rests.
The step sizes and the waiting times are drawn from a \gls{pdf}, called the \textit{jump PDF}, $\psi(x,t)$.
Various choices of the form of $\psi(x,t)$ lead to different situations.
For example, if the step sizes and the waiting times are independent random variables, the \gls{ctrw} is decoupled and
the jump \gls{pdf} can be written as $\psi(x,t)=\lambda(x) w(t)$, where $\lambda(x)$ and $w(t)$ are the step size
\gls{pdf} and the waiting time \gls{pdf}, respectively.
In the case of a coupled jump \gls{pdf}, a jump of a certain length involves a certain duration.
This is, for example, the case of the L\'evy-walk.

The \gls{ctrw} can be described by the equation\cite{Klafter1987}
\be
\eta(x,t)=\int_{-\infty}^{+\infty}\diff x'\int_0^{+\infty}\diff t'\eta(x',t')\psi(x'-x,t-t')+\delta(x)\delta(t),
\ee
which links the \gls{pdf} $\eta(x,t)$ of arriving at position $x$ at time $t$ with the \gls{pdf} $\eta(x',t')$ of being
arrived at position $x'$ at time $t'$ with a delta Dirac initial condition.

For the decoupled case, the \gls{pdf} of the density of walkers is therefore given by
\be\label{eq:ctrw}
n(x,t)=\int_0^{t}\diff t'\eta(x,t')\Psi(t-t')\diff t',
\ee
where $\Psi(t)=1-\int_0^t w(t')\diff t'$ is the probability for a walker of making no jumps between the time interval
$(0,t)$.

By taking the space Fourier transform and time Laplace transform\footnote{
In the following we us the notation $\hat{f}$ for both the Fourier and the Laplace transforms of the function $f$.
The difference between the two transforms is indicated by their conjugate variables: $x\xrightarrow{FT} k$ and
$t\xrightarrow{LT} s$.
}
of eq. (\ref{eq:ctrw}), one finds the
\textit{Montroll-Weiss} equation
\be\label{eq:montroll}
\hat{n}(k,s)=\frac{1-\hat{w}(s)}{s(1-\hat{w}(s)\hat{\lambda}(k))}.
\ee

Starting from this equation, the fluid limit, meaning that all details of the \gls{ctrw} that are irrelevant at
very large temporal and spatial scales are neglected, is usually sought to find a equation describing the
temporal evolution of the density of random walker.

For example, classical diffusion is found by taking a Gaussian distribution of step sizes and a Poissonian wating time
\gls{pdf}.
In fact, the fluid limit is always found to be the classical diffusion, as long as the characteristic waiting time
\be
\tau=\int_0^{+\infty}t w(t)\diff t
\ee
and the variance of step sizes
\be
\Delta^2=\int_{-\infty}^{+\infty}x^2 \lambda(x)\diff x
\ee
are finite\cite{Metzler2000}. In this case the diffusion coefficient is given by $D=\Delta^2/\tau$.

To describe \ndt, $\lambda$ and $w$ have to be chosen such that $\tau$
or $\Delta^2$, or both, diverge.
A natural choice is to draw both functions in the family of stable distributions (with certain restrictions) since they
are the limit of sums of random variables\cite{Sanchez2005}.
The non-Gaussian property will arise from a divergent $\Delta^2$ and the non-Markovianity from a divergent $\tau$.
In the case where both characteristic scales diverge, we can choose a strictly symmetric stable distribution with
characteristic exponent $\alpha<2$ for the step sizes distribution and a one-sided stable distribution with
characteristic exponent $\beta<1$ for the waiting time distribution (see appendix \ref{sec:stable_dist}):
\be
\hat{\lambda}(k)=e^{-\sigma^\alpha|k|^\alpha}\simeq1-\sigma^\alpha|k|^\alpha\text{\,\,\, for \,\,\,}k\rightarrow0
\text{\,\,\, with \,\,\,} \alpha<2
\ee
and
\be
\hat{w}(s)=e^{-\mu^\beta s^\beta}\simeq1-\mu^\beta s^\beta\text{\,\,\, for \,\,\,}s\rightarrow0  \text{\,\,\, with
\,\,\,} \beta<1,
\ee
with the following asymptotic behavior
\be
\lambda(x)\sim x^{-(\alpha+1)},\text{\,\,\, when \,\,\,} |x|\rightarrow\infty 
\ee
and
\be
w(t)\sim x^{-(\beta+1)},\text{\,\,\, when \,\,\,} t\rightarrow\infty.
\ee
Using those expression in eq. (\ref{eq:montroll}) and keeping terms up to first order, one finds
\be\label{eq:FL_crtw}
\hat{n}(k,s)=\frac{s^{\beta-1}}{s^\beta+\frac{\sigma^\alpha}{\mu^\beta}|k|^\alpha},
\ee
which can be rearranged as
\be
s^\beta\hat{n}(k,s)-s^{\beta-1}=-\frac{\sigma^\alpha}{\mu^\beta}|k|^\alpha\hat{n}(k,s).
\ee
If $\alpha=2$ and $\beta=1$, one recognizes the Laplace transform of the first order time derivative of $\hat{n}(k,t)$
on the \gls{lhs} and the Fourier transform of the second order space derivative of $\hat{n}(x,s)$ on the \gls{rhs}.
Therefore, in this case, we recover the classical diffusion equation (eq. (\ref{eq:diffusion})) with $D=\sigma^2/\mu$.

In the case were $\alpha<2$ and $\beta<1$, the \gls{lhs} and \gls{rhs} correspond to generalizations of the
differential operator to fractional orders (see appendix \ref{sec:fract_ope}) 
\be\label{eq:space_time_fractional_diffusion}
\frac{\uppartial^\beta}{\uppartial t^\beta}n(x,t)=D_{\alpha,\beta}\frac{\uppartial^\alpha}{\uppartial |x|^\alpha}n(x,t),
\ee
where $D_{\alpha,\beta}=\sigma^\alpha/\mu^\beta\,[\si{m^\alpha/s^\beta}]$ is a generalized diffusion coefficient.
This equation is called the \textit{space and time fractional diffusion equation} and describes the time evolution of
the fluid limit of a \gls{ctrw} with long-term memory (non-Markovian) and long-range spatial correlations
(non-Gaussian).
When $\alpha\rightarrow2$ and $\beta\rightarrow1$, the classical diffusion equation (eq. (\ref{eq:diffusion})) is 
recovered.
We call $\alpha$ the \textit{spatial transport exponent} and $\beta$ the \textit{temporal transport exponent}.
The time derivative operator is the \textit{Caputo fractional derivative} and the space derivative is the \textit{Riesz
fractional derivative} (appendix \ref{sec:fract_ope}).

By using the scaling properties of the Fourier and Laplace transform
\be
\mathcal{F}[f(ax)]=|a|^{-1}\hat{f}(k/a) \text{\,\,\, and \,\,\,} \mathcal{L}[f(bu)]=b^{-1}\hat{f}(s/b),\,\,\, b>0,
\ee
on eq. (\ref{eq:FL_crtw}), the following scaling property of the propagator of 
eq. (\ref{eq:space_time_fractional_diffusion}) is
inferred
\be\label{eq:propagator_scaling}
G_{\alpha,\beta}(x,t)=t^{-H}K_{\alpha,\beta}\left(\frac{x}{t^H}\right),
\ee
where $\hurst=\beta/\alpha$ is the \textit{self-similarity index} also called \textit{Hurst exponent}\cite{Saenko2009},
$K_{\alpha,\beta}$ is called the \textit{reduced Green function} and $\frac{x}{t^H}$ is the
\textit{similarity variable}\cite{Mainardi2001}.
This scaling implies that the stochastic process associated with the propagator is \textit{self-similar} with index
$\hurst=\beta/\alpha$ (or \hurst-self similar).
Indeed, if we rescale the time by a factor $\lambda>0$, we find
\be
G_{\alpha,\beta}(x,\lambda t)=\lambda^{-H}t^{-H}K_{\alpha,\beta}\left(\frac{x}{t^H\lambda^H}\right)\propto
G_{\alpha,\beta}(x\lambda^{-H},t),
\ee
implying that the motion is invariant under the following transformation
\be\label{eq:H-ss}
(x,t)\mapsto(\lambda^Hx,\lambda t).
\ee
The self-similarity of the process leads then to the following scaling of the moments of the distribution with time
\be\label{eq:s-moment-scaling}
\left<|x|^s\right>\propto t^{sH}.
\ee
Indeed, if we assume $\left<|x|^s\right>\propto t^K$, we have
\be
\left<|\lambda^Hx|^s\right>\propto (\lambda t)^K \,\,\,\,\Rightarrow\,\,\,
\lambda^{sH}\left<|x|^s\right>\propto \lambda^K t^K\,\,\,\,\Rightarrow\,\,\,
K=sH.
\ee
By identifying eq. (\ref{eq:msd_gamma}) with eq. (\ref{eq:s-moment-scaling}) we see that $\gamma=2H$.
The transport is therefore superdiffusive if $2\beta>\alpha$ and subdiffusive if $2\beta<\alpha$.
For the special case $2\beta=\alpha,\beta\neq1,\alpha\neq2$ the transport is called
\textit{quasidiffusive}\cite{Brockmann2006} and exhibits the same scaling as the classical diffusion despite the crucial
difference of a non-Gaussian \gls{pdf} and non-Markovian time increments.
We would like to point out that only verifying the time dependence of the variance of displacements is not
sufficient to
conclude that a transport process is diffusive.
Whenever possible, the temporal evolution of higher moments, or of the full distribution, should be examined to
determine the presence of non-Gaussian features.
Evaluating moments of the distribution can be delicate since the integrals that define them do not always converge.
For example, the $s$-order moments of a $\alpha$-stable distributions converges only for
$s<\alpha$\cite{Samorodnitsky1994}.
However, in practice, the temporal scaling can be recovered by taking ``truncated'' moments or by computing moments of
fractional
order smaller than $\alpha$\cite{Metzler2000}.

\subsubsection{L\'evy flight}

\begin{figure}[!tb]
\centering
\includegraphics[width=0.7\linewidth]{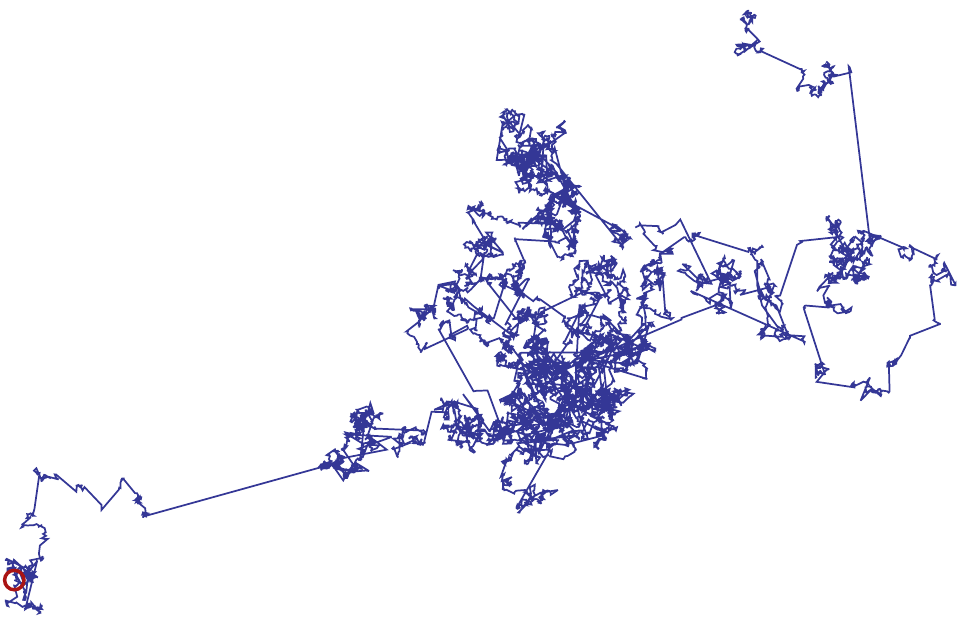}
\caption{Example of a L\'evy flight trajectory with $\alpha=1.5$.
Contrary to the Brownian motion (fig. \ref{fig:PerrinPlot}), arbitrary long steps can arise, on all scale length,
leading
to the clustering nature of the L\'evy flight.} 
\label{fig:levy_flight}
\end{figure}

The L\'evy flight is a particular case of the \gls{ctrw} with a finite characteristic waiting time, $\tau<\infty$,  but
step size distribution given by a symmetric stable distribution with diverging variance, $\Delta^2\rightarrow\infty$.
The trajectories of L\'evy flights have been shown to model the foraging motions of many living
organisms\cite{Ayala-Orozco2004,Reynolds2009}.
Mandelbrot also used this model to simulate the fractal galaxy distribution in the Universe\cite{Mandelbrot1983fractal}.
In fact, a fractal dimension $D=\alpha$ can be assigned to the trajectories.
An example of a L\'evy flight trajectory is shown in fig. \ref{fig:levy_flight}.

The L\'evy flight can be modeled by taking a Poissonian distribution for the waiting time
\gls{pdf} $w(t)=\tau^{-1}e^{-t/\tau}$, with Laplace transform $\hat{w}(s)=(1+s\tau)^{-1}\simeq1-\tau s$, for
$s\rightarrow0$.
Using this in eq. (\ref{eq:montroll}), we find
\be\label{eq:levy_flight_FL}
\hat{G}_\alpha(k,s)=\frac{1}{s+\frac{\sigma^\alpha}{\tau}|k|^\alpha},
\ee
which, upon Fourier-Laplace inversion, shows that the propagator of the L\'evy flight is a symmetric stable distribution
\be\label{eq:levy_flight_propagator}
{G}(x,t)=t^{-H}L_{\alpha,\sigma}\left(\frac{x}{t^H}\right),
\ee
with a self-similar index $\hurst=1/\alpha$. Rearranging the terms of eq. (\ref{eq:levy_flight_FL}) and Fourier-Laplace
inverting
it, we find the \textit{space fractional diffusion equation}
\be\label{eq:levy_flight_equ}
\frac{\uppartial}{\uppartial t}n(x,t)=D_{\alpha}\frac{\uppartial^\alpha}{\uppartial |x|^\alpha}n(x,t)
\ee
describing the L\'evy flight.

The L\'evy flight results in a superdiffusive process (with the exception of the case $\alpha=2$) with a diverging
mean-square displacement $\left<x^2\right>\rightarrow\infty$ for $\alpha<2$.

The presence of arbitrary long jumps without any restriction on the step duration leads to rather
unphysical situations\cite{Metzler2000,Metzler2004}.
One way of solving this is introduced in the L\'evy walk model, which is often more appropriate to
describe physical systems.

\subsubsection{L\'evy walk}
\label{sec:levy_walk}
Similarly to the L\'evy flight, the L\'evy walk model maintains a diverging variance of distribution of step sizes.
However, a coupling between the step sizes and the step duration is included in the jump pdf such 
that\cite{Klafter1987,Gustafson2012levy}
\be
\psi(x,t)=\lambda(x)\delta(|x|-vt^\nu),
\ee
where $\lambda(x)\rightarrow|x|^{-\mu}$ as $|x|\rightarrow\infty$ and $v$ is a generalized velocity which penalizes
long jumps such that the variance is finite\cite{Metzler2004}.
Depending on the values of the two exponent $\mu$ and $\nu$ the transport can be either superdiffusive or subdiffusive.
Due to the coupled form of the jump pdf, the derivation of a transport equation describing the evolution of the
\gls{pdf}
has only been achieved recently in the case $\nu=1$, by using a fractional version of the material
derivative\cite{Sokolov2003}.

\subsection{Langevin approach}

A microscopic description of Brownian motion equivalent to the one of Einstein presented in sec. 
\ref{sec:classical_diff}
was introduced by Langevin in 1908\cite{Langevin1908}
An uncorrelated Gaussian noise, representing the random force due to the interaction with the fluid molecules, is
used in the equation of motion of a test particle.
The equation of motion becomes a stochastic equation, whose average motion shows the same diffusive scaling.
\Gls{fbm} introduces long-range temporal dependence in the Gaussian noise, which can lead to a non-linear scaling of the
positional variance.
On the other hand, non-Gaussian statistics can be introduced by choosing a non-Gaussian noise.
For example, stable L\'{e}vy motion\cite{Samorodnitsky1994} replaces the Gaussian noise with a random noise
distributed according to a L\'{e}vy stable distribution with heavy-tails (appendix \ref{sec:stable_dist}).

The classical Langevin equation is written\cite{Langevin1908}
\be\label{eq:Langevin}
m\ddot{x}(t)=-m\gamma\dot{x}(t)+\xi(t),
\ee
where $m$ is the mass of the test particle, $\gamma$ is the friction coefficient and $\xi(t)$ is the random force due
to the random collisions with the surrounding particles.
In the case of the Brownian motion, $\xi(t)$ is a a white noise, i.e. a Gaussian noise with an infinitely short
correlation time: $\left<\xi(t_1)\xi(t_2)\right>=C\delta(t_1-t_2)$, where $C$ is a constant. 
In eq. \ref{eq:Langevin}, the forces acting on the particle are separated in two groups, the macroscopic, slowly varying
ones, represented by the dissipative force $-m\gamma v$, and the microscopic, rapidly varying ones, represented by the
fluctuating force $\xi(t)$.

If the time scale of the particle motion is comparable to the time scale of the collisions, the assumption of a
white noise and a constant friction have to be abandoned. 
This leads to the \acrfull{gle}\cite{Kubo1966}
\be\label{eq:generalized_Langevin}
\ddot{x}(t)=-\int_0^t\beta(t-t')\dot{x}(t')\diff t'+\xi(t),
\ee
where $m=1$ is used for simplicity.
Here, $\beta(t)$ is the memory kernel and $\xi(t)$ is the random force which is zero-centered and stationary, i.e.
$\left<\xi(t_1)\xi(t_2)\right>=C(|t_1-t_2|)=C(\tau)$.
The fluctuation-dissipation theorem\cite{Nyquist1928fluctdiss,Callen1951fluctdiss,Kubo1966,Porr1996} states that the
dissipation is the macroscopic manifestation of the disordering effect of the fluctuations and relates the
correlation function  of the random forces $C(t)$ with $\beta(t)$ by
\be\label{eq:fluct-dissp}
k_BT\beta(t)=C(t).
\ee
Assuming $x(0)=0$, $v(0)=v_0$ and Laplace transforming eq. (\ref{eq:generalized_Langevin}), we find
\be
\hat{x}(s)=\frac{v_0+\hat{\beta}(s)+\hat{\xi}(s)}{s(s+\hat{\beta}(s))},
\ee
where $\hat{\beta}(s)$ and $\hat{\xi}(s)$ are the Laplace transforms of $\beta(t)$ and $\xi(t)$.
Upon Laplace inversion, one finds the equation of the particle position
\be
x(t)=v_0H(t)+\int_0^tH(t-t')\xi(t')\diff t',
\ee
where $H(t)$ is the relaxation function\cite{Porr1996} defined by its Laplace transform
\be
\hat{H}(s)=\frac{1}{s(s+\hat{\beta}(s))}.
\ee
We note that, in accordance with the classical Langevin equation, if $\beta(t)=\gamma=\text{cste.}$, the relaxation
function, $H(t)=\frac{1}{\gamma}(1-e^{-\gamma t})$, is exponentially decaying and the position is given by
\be\label{eq:classical_langevin}
x(t)=\frac{v_0}{\gamma}(1-e^{-\gamma t})+\frac{1}{\gamma}\int_0^t(1-e^{-\gamma (t-t')})\xi(t')\diff t'.
\ee
For $t\rightarrow0$ the ballistic motion $x(t)=v_0 t$ is recovered and for $t\gg\gamma^{-1}$,
\be\label{eq:ordinaryBorwnian motion}
x(t)=\frac{v_0}{\gamma}+\frac{1}{\gamma}\int_0^t\xi(t')\diff t'.
\ee
When $v_0=0$, eq. (\ref{eq:ordinaryBorwnian motion}) is referred to as \acrfull{obm}.

If we take $\xi$ as a white noise and $\left<v_0\right>=0$, the variance of the displacement is given by
\begin{align}
\left<x^2(t)\right>&=\frac{1}{\gamma^2}\left<\int_0^t\xi(t-t_1)\diff t_1\int_0^t\xi(t-t_2)\diff t_2\right> \notag\\
    &=\frac{1}{\gamma^2}\left<\int_0^t \diff t_1 \int_0^t \diff t_2 \xi(t-t_1) \xi(t-t_2)\right>\notag\\
 &= \frac{1}{\gamma^2}\int_0^t \diff t_1 \int_0^t \diff t_2 \left<\xi(t-t_1) \xi(t-t_2)\right>\notag\\
 &= \frac{1}{\gamma^2}\int_0^t \diff t_1 \int_0^t \diff t_2 C\delta(t_2-t_1)\notag\\
 &= \frac{1}{\gamma^2}\int_0^t C \diff t_1 = \frac{C}{\gamma^2}t=\frac{k_B T}{\gamma}t,
\end{align}
where we used eq. (\ref{eq:fluct-dissp}).
Thus, we recover the linear temporal scaling of the mean-squared displacement of the classical diffusion
(eq. (\ref{eq:variance_diff_lin})).

\subsubsection{Fractional Brownian motion}

\Acrfull{fbm} was proposed by Mandelbrot and Van Ness in 1968\cite{Mandelbrot1968} to
model the variations of cumulated water flows in the great lakes of the Nile river basin observed by
Hurst\cite{Hurst1951}.
Hurst studied the record of river level and other physical quantities such as rainfall, temperature, pressure, the
growth of tree rings, sunspot numbers and wheat prices.
He found that the range of those records, rescaled by their standard deviation, is proportional to $t^H$, where $t$ is
the time and $1/2<\hurst<1$ is, ever since, called the Hurst exponent.
Since then, it has found a wide range of applications in systems showing long time interdependence.

Slightly different representations exists in the literature, here we use the following\cite{Calvo2008}
\be
x_H(t)=\frac{1}{\Gamfnct(H+1/2)}\int_0^t(t-t')^{H-1/2}\xi(t')\diff t',
\ee
where $x_H(t)$ represents the position of a particle experiencing \gls{fbm}, $\xi(t)$ is a Gaussian uncorrelated
noise, $H\in(0,1]$ and $\Gamfnct(\cdot)$ is the gamma function.
\Gls{fbm} is constructed as a moving averaged of the \gls{obm} (eq. (\ref{eq:ordinaryBorwnian motion})), in which
past increments are weighted by the power law kernel $(t-t')^{H-1/2}$.
It has a zero mean $\left<x_H(t)\right>=0$ for $\hurst<1$.
From its definition and the fact that the Gaussian noise is self-similar with exponent $1/2$, it follows that
$x_H(t)$ is \hurst-self similar (eq. (\ref{eq:H-ss})) and that it has stationary increments,
$x_H(t)-x_H(s)=x_H(t-s)$\cite{Samorodnitsky1994}.
Using these two properties, we can show that the correlation function is
\begin{align}
\left<x_H(t)x_H(s)\right>&=K_H\frac{1}{2}\left\{
\left<x_H^2(t)\right>+\left<x_H^2(s)\right>-\left<(x_H(t)-x_H(s))^2\right>\right\}\notag\\
&=K_H\frac{1}{2}\left\{\left<x_H^2(t)\right>+\left<x_H^2(s)\right>-\left<x_H^2(t-s)\right>\right\}\notag\\
&=K_H\frac{1}{2}\left\{t^{2H}+s^{2H}-|t-s|^{2H}\right\},
\label{eq:fBm_corr}
\end{align}
where $K_H$ is a positive constant and where we recall that for a \hurst-self similar process, the variance scales
with time as
\be
\left<x_H^2(t)\right>\propto t^{2H}.
\ee
We note that \gls{fbm} is subdiffusive for $0<\hurst<1/2$, superdiffusive for $1/2<\hurst<1$, ballistic for $\hurst=1$
and correspond
to \gls{obm} for $\hurst=1/2$.

The increments of the \gls{fbm}, $\xi_H(t)$, is a stationary Gaussian process known as \acrfull{fgn} and defined as
\be
x_H(t)=\int_0^t\xi_H(t')\diff t'.
\ee
The correlation function, $C_H(t)$, of $\xi_H(t)$ is given by the derivative of equation \eqref{eq:fBm_corr} with
respect to $t$ and $s$
\be
C_H(|t-s|)=\left<\xi_H(t)\xi_H(s)\right>=2K_H H (2H-1)|t-s|^{2H-2}+2K_H H|t-s|^{2H-1}\delta(t-s).
\ee
We note that $C_H(|t-s|)$ behaves as a power law for $\tau=|t-s|\rightarrow\infty$ and recovers the ordinary Brownian
behavior, $C_H(|t-s|)=K_{1/2}\delta(t-s)$, for $\hurst=1/2$.

\begin{figure}[!tb]
\centering
\includegraphics[width=0.6\linewidth]{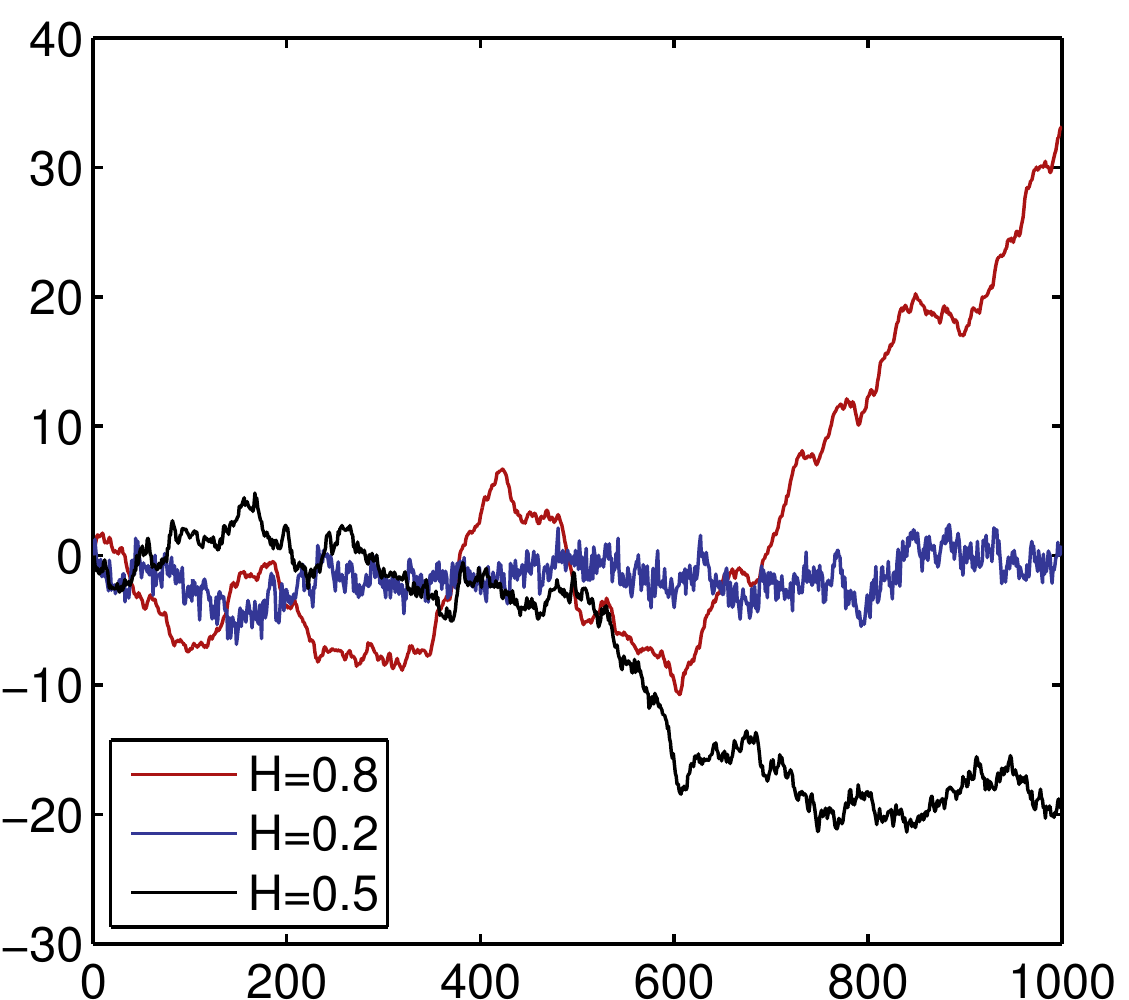}
\caption{Examples of realizations of the \acrfull{fbm} for different values of the self-similar index \hurst.
For $\hurst=0.8$, the increments of the trajectory are positively correlated which results in a persistent motion.
The \gls{obm} with uncorrelated increments is retrieved for $\hurst=0.5$ and for $\hurst=0.2$, the increments of
the trajectory are anti-correlated which results in a anti-persistent motion.
} 
\label{fig:fBm}
\end{figure}

The function $C_H(\tau)$ tends to zero for $\tau=|t-s|\rightarrow\infty$ for $0<\hurst<1$, but when $1/2<\hurst<1$,
$\xi_H(t)$
exhibits long-range dependence, i.e. $C_H(\tau)$ tends to zero so slowly that $\int_0^\infty C_H(\tau)d\tau=\infty$.
It is said to be \textit{correlated}.
For $0<\hurst<1/2$, there is no long-range dependence, but the coefficient $(2H-1)$ is negative\cite{Samorodnitsky1994}.
In this case the $\xi_H(t)$ is said to be \textit{anti-correlated}.
Figure \ref{fig:fBm} shows three examples of \gls{fbm} trajectories.

In the framework of the \gls{gle}, it is possible to find the \gls{fbm} as a solution by using a random force with
long-range correlations, namely with a power-law correlation function.
The memory kernel, $\beta(t)$, is then found with the fluctuation-dissipation theorem (eq. (\ref{eq:fluct-dissp})) and,
consequently, also have a power-law form.
When the random force is chosen to be the \gls{fgn}, the \gls{gle} can be written as a fractional differential
equation\cite{Jeon2010}, however, the solution of this equation is limited to the subdiffusive and diffusive case.
From the physical point of view, the superdiffusive case is found only when the random force is ``external'', meaning
that the fluctuation-dissipation theorem does not hold and that the driving noise and the dissipation may have different
origins, which may be the case in nonequilibrium systems\cite{Porr1996}.

It is possible to find the propagator of the \gls{fbm} by using the method of path integrals\cite{Calvo2008}, borrowed
from quantum mechanics (note that Shr\"odinger's equation resembles a diffusion equation with an imaginary diffusion
coefficient).
The propagator, given by
\be\label{eq:fBm_propagator}
G_H(x,t)=\sqrt{\frac{H}{\pi}}\frac{\Gamfnct(H+1/2)}{t^H}\exp\left[-H \Gamfnct^2(H+1/2)\frac{x^2}{t^{2H}}\right],
\ee
is a Gaussian function with a variance proportional to $t^{2H}$.
It has the same form than the propagator of the \gls{obm} (eq. (\ref{eq:ordinaryBorwnian_propagator})) but
with a ``stretched'' time $t^{2H}$.
The transport equation of the \gls{fbm} is easily found from the Fourier transform of eq. (\ref{eq:fBm_propagator}) to 
be
\be\label{eq:fBm_diffeq}
\frac{\uppartial}{\uppartial t}n(x,t)=D_\beta t^{\beta-1}\frac{\uppartial^2}{\uppartial x^2}n(x,t),
\ee
where $0<\beta=2\hurst<2$ and $D_\beta=(2\Gamfnct^2(H+1/2))^{-1}$ is a stretched diffusion coefficient of dimensions
$[m^2/s^\beta]$. 
This equation is called the \textit{stretched time diffusion equation}\cite{Mainardi2010}.
By using the rule
\be
\frac{\uppartial}{\uppartial t^\beta}=\beta t^{\beta-1} \frac{\uppartial}{\uppartial t},
\ee
it can be interpreted as the result of the classical diffusion equation with a stretched time.

We note that the Langevin approach and the \gls{ctrw} approach are not equivalent in the non-Markovian case.
Equation (\ref{eq:fBm_diffeq}) is local in time, whereas eq. (\ref{eq:space_time_fractional_diffusion}) with 
$\alpha=2,\beta<1$, the
\textit{time-fractional diffusion equation}, is not. 
In the \gls{fbm} case, the non-Markovian character is provided by a time dependent diffusivity $D=D_0t^{\beta-1}$.
Moreover, The solution of eq. (\ref{eq:fBm_diffeq}) is Gaussian, while the solution of the time-fractional diffusion 
equation
is not; it is given by the transcendental functions known as the  M-Wright function which tends to the Gaussian function
for $\beta=1$\cite{Mainardi2010}.

%

\subsubsection{Fractional L\'evy motion}
\label{sec:fLm}

Here, we discuss the \acrfull{flm}, which is a generalization of the \gls{fbm}, including both long-range
temporal dependence and non-Gaussian statistics.

The stochastic equation defining the \gls{flm} process is\cite{Laskin2002,Calvo2009}
\begin{equation}
 x_{\alflm,H}(t)=\frac{1}{\Gamfnct\left(H-1/\alflm+1\right)}\int_0^t(t-t')^{H-1/\alflm}\xi_{\alflm,\sigma}(t')\diff 
t'\mbox{,}
\end{equation}
where $\xi_{\alflm,\sigma}(t)$ is an uncorrelated noise distributed according to a L\'{e}vy symmetric,
strictly stable distribution, with index of stability $\alflm$, ($0<\alflm\leq2$) and scale parameter $\sigma$.
From the properties of $\alflm$-stable random variable, we have $x_{\alflm,H}(\lambda t)=\lambda^{H} x_{\alflm,H}(t)$,
with
$\hurst=\beflm/\alflm$.
Therefore, the \gls{flm} belongs to the important family of \hurst-self similar process with stationary increments
(also abbreviated \textit{H-sssi}), like the \gls{fbm}.
Consequently, the moments of $x_{\alflm,H}(t)$ exhibit the desired general {non-classical} feature
\be
\left\langle |x_{\alflm,H}(t)|^s \right\rangle \propto t^{sH},
\ee
where $0<s<\alflm$, to ensure convergence of the moments.
For a non-degenerated process, the values of \hurst are restricted to\cite{Samorodnitsky1994}
\begin{equation}
\left\lbrace
\begin{array}{lll}\label{eq:rang_of_H}
 0<H\leq 1/\alflm & \mbox{ if } & \alflm<1 \mbox{,}\\
 0<H\leq1         & \mbox{ if } & \alflm\geq1 \mbox{.}
\end{array}
\right.
\end{equation}
The long-term memory is engendered by the convolution with the power-law kernel and the non-Gaussian statistics
by the L\'{e}vy noise. The fLm generalizes the fractional Brownian motion (fBm)\cite{Mandelbrot1968}. Indeed,
for $\alflm=2$, the noise has a Gaussian distribution and the process is the fBm.
When $\hurst=1/\alflm$ the process is time-uncorrelated and when $\hurst<1/\alflm$ or $\hurst>1/\alflm$ the process
exhibits negative
or positive time correlations, respectively. Therefore, for $\alflm=2$ and $\hurst=1/2$, one recovers the \gls{obm}
corresponding to classical diffusion (eq. (\ref{eq:ordinaryBorwnian motion})).

\begin{figure}[!tb]
\centering
\includegraphics[width=\linewidth]{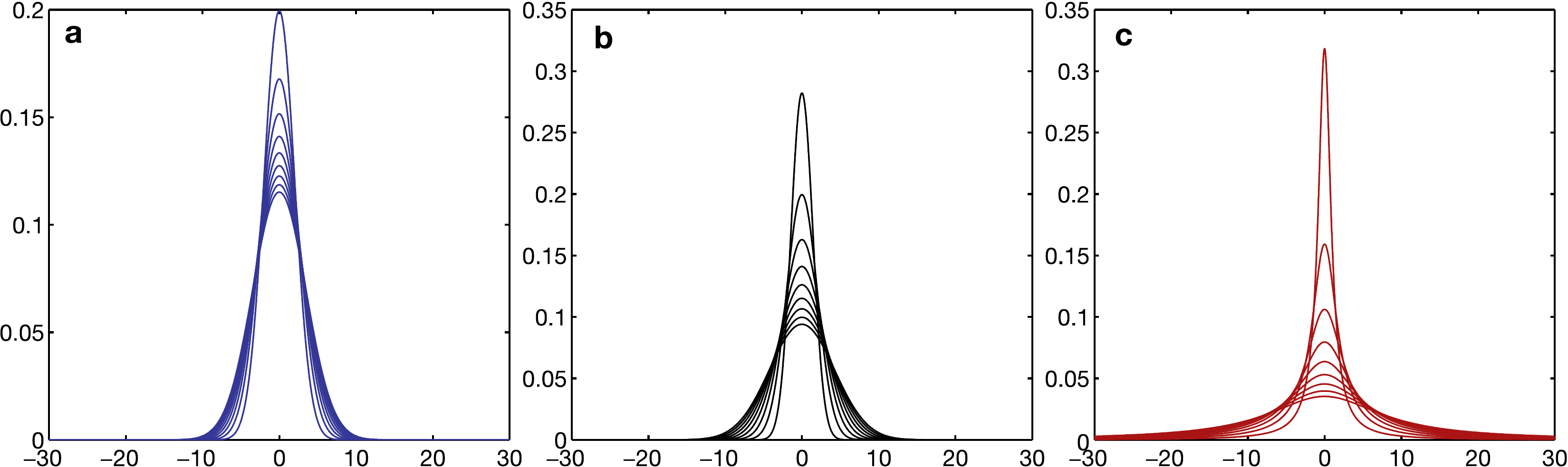}
\caption{Examples of propagators for different parameters of the \gls{flm} for times $t_i=1,2,...,9$ and effective
diffusivity $K=1$.
\textbf{a}: Subdiffusive case with $\alflm=2$ (Gaussian) and $\beflm=0.5$ (anti-correlated) ($\hurst=0.25$).
\textbf{b}: Diffusive case with $\alflm=2$ (Gaussian) and $\beflm=1$ (uncorrelated) ($\hurst=0.5$).
\textbf{c}: Superdiffusive case with $\alflm=1$ (L\'evy stable) and $\beflm=1$ (uncorrelated) ($\hurst=1$).
} 
\label{fig:fLm_propa}
\end{figure}

Using path integrals, Calvo, S\'{a}nchez and Carreras have shown that the transport equation of the \gls{flm} process is
a \textit{space-fractional diffusion equation} with time dependent diffusivity\cite{Calvo2009}
\begin{equation}\label{eq:flm_frac_diff}
 \frac{\uppartial}{\uppartial t}n(x,t) = K t^{\beflm -1} \frac{\uppartial^\alpha }{\uppartial \left|
x\right|^\alpha } n(x,t)\mbox{.}
\end{equation}
Here $n(x,t)$ is the density of particles, $K$ is an effective diffusion coefficient and $\alflm$ and $\beflm$ are
the space and time transport exponents, respectively, with $\hurst=\beflm/\alflm$. The space derivative of order
$\alflm$
is
the 
Riesz fractional differential operator\cite{Podlubny1999} (appendix \ref{sec:fract_ope}).
The restriction on the range of permissible values for \hurst (eq. (\ref{eq:rang_of_H})) translates for $\beflm$ as
\begin{equation}
\left\lbrace
\begin{array}{lll}
 0<\beflm \leq 1 & \mbox{ if } & \alflm<1 \mbox{,}\\
 0<\beflm \leq \alflm         & \mbox{ if } & \alflm\geq1 \mbox{.}
\end{array}
\right.
\end{equation}
When $\beflm<1$ or $\beflm>1$ the process is negatively or positively time correlated, $\beflm=1$ corresponding to an 
uncorrelated process.

The propagator of eq. (\ref{eq:flm_frac_diff}) is a L\'{e}vy distribution which depends on $x/t^H$\cite{Calvo2009}
\begin{equation}\label{eq:flm_propagator}
 G_{\alflm,\beflm}(x,t)=
 \frac{C(\alflm,\beflm)}{t^{\beflm/\alflm}}L_{\alflm,\sigma}\left[C(\alflm,\beflm)\frac{x}{t^{\beflm/\alflm}}
\right],
\end{equation}
where $C(\alflm,\beflm)=\beflm^{1/\alflm}\Gamfnct\left(\frac{\beflm-1}{\alflm}+1\right)$ and $\sigma=K^{1/\alflm}
\Gamfnct\left(\frac{\beflm-1}{\alflm}+1\right)$. 

\begin{figure}[!tb]
\centering
\includegraphics[width=0.7\linewidth]{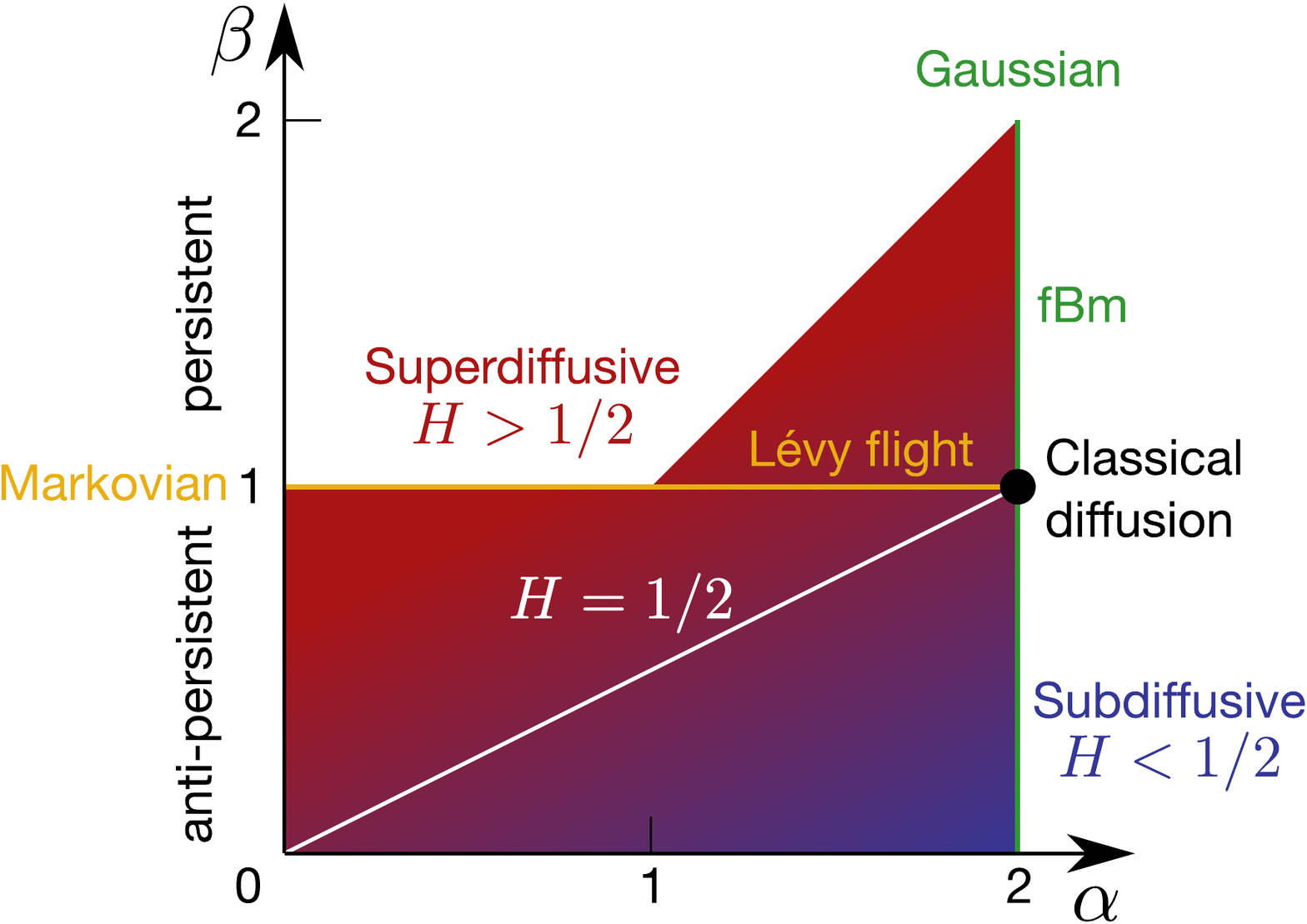}
\caption{
Parameter space for the \gls{flm}.
The values of $\hurst=\beflm/\alflm$ lie in the shaded region and depending on the values of the spatial transport
exponent,
$\alflm$, and the temporal transport exponent, $\beflm$, the transport can be superdiffusive ($\hurst>1/2$) or
subdiffusive
($\hurst<1/2$). Diffusive transport is found for $\hurst=1/2$ (white line), the classical diffusion corresponding to the
case
$\beflm=1$ and $\alflm=2$ (black dot).
Gaussian transport is found for $\alflm=2$ and Markovian transport for $\beflm=1$.
For $\beflm>1$ the motion is persistent and for $\beflm<1$ it is anti-persistent.
The \gls{fbm} (green line) is found for $\alflm=2$ and $0<\beflm<2$ and the L\'evy flight (yellow line) for $\alflm<2$
and $\beflm=1$.
A similar figure can be drawn for the \gls{ctrw}, but with the difference that the non-Markovianity is not due to
persistence or anti-persistence but to the non-locality in time and, in this case, $\beflm\leq1$.
} 
\label{fig:alpha_beta_parameterspace}
\end{figure}

Again, for $\alflm=2$ the space-fractional derivative becomes a second order derivative and the propagator is a
Gaussian corresponding to the case of the \gls{fbm} (eq. (\ref{eq:fBm_propagator}) and eq. (\ref{eq:fBm_diffeq})).
For $\alflm=2$ and $\beflm=1$ ($\hurst=1/2$), eq. (\ref{eq:flm_frac_diff}) becomes the classical diffusion equation and
the propagator has the well known form of a Gaussian with variance growing linearly with time (eq. (\ref{eq:diffusion}) 
and eq. (\ref{eq:ordinaryBorwnian_propagator})).
Finally, for $\beflm=1$ and $\alflm<2$, eq. (\ref{eq:flm_propagator}) and eq. (\ref{eq:flm_frac_diff}) are Markovian
and correspond to the propagator and transport equation of the L\'evy flight (eq. (\ref{eq:levy_flight_propagator}) and
eq. (\ref{eq:levy_flight_equ})).
Examples of \gls{flm} propagators are shown in eq. (\ref{fig:fLm_propa}) and the different transport regimes of the
\gls{flm}, as a function of $\alflm$ and $\beflm$ are summarized in fig. \ref{fig:alpha_beta_parameterspace}.
 
We note here that a generalization of the model of the \gls{flm} to include asymmetric probability density functions as 
recently been derived and applied to the case of suprathermal ion transport in turbulent 
plasmas\cite{Bovet2014nf,Bovet2015phd}.

\section*{Acknowledgments}
The author wish to acknowledge the precious discussions, remarks and the proofreading of Ambrogio Fasoli, Ivo Furno, 
Paolo Ricci, Kyle Gustafson, Gaetano Zimbardo, Fulvio Zonca and Frederick Skiff.
This work was supported in part by the Swiss National Science Foundation.

\appendix
\section{Stable distributions}

\label{sec:stable_dist}

We recall here the main properties of $\alpha$-stable random variables and distributions
(also called L\'{e}vy $\alpha$-stable distributions), although we refer the reader to reference \cite{Samorodnitsky1994}
for a detailed monograph.
The generalized central limit theorem states that stable distributions are the limiting distributions of normalized sums
of independent, identically distributed random variables.
The Gaussian distribution is a particular case that corresponds to the case where the random variables have a finite
variance.
Stable distributions are especially interesting because, with the exception of the Gaussian distribution, they allow
heavy-tails and non-zero skewness.
Their probability densities exist and are continuous, but their expressions in closed form exist only for a few
particular cases (Gaussian, Cauchy and L\'{e}vy distributions) and they are usually described by their characteristic
function, i.e. their Fourier transform.

A random variable $X$ is said to be stable if for any $A,B>0$ there is $C>0$ and $D\in\mathbb{R}$ 
such that\cite{Samorodnitsky1994}
\be
AX+BX=CX+D.
\ee
Moreover, for any stable random variable there is a number $\alpha \in (0,2]$ such that
\be
C^\alpha=A^\alpha+B^\alpha.
\ee
The Gaussian distribution corresponds to the case $\alpha=2$.

Different parameterizations of their characteristic function are possible.
We adopt here the following parameterization for the stable distribution $L^\theta_{\alpha,\sigma,\mu}$
(close to parameterization C in \cite{Zolotarev1986}):
\begin{equation}
 \mathcal{F}[L^\theta_{\alpha,\sigma,\mu}(x)](k)=\exp\left\{-\sigma^\alpha|k|^\alpha e^{i
\sign(k)\frac{\theta\pi}{2}}+ik\mu\right\}.
\end{equation}
Here $\alpha \in (0,2]$ is the \textit{index of stability} or \textit{characteristic exponent},
$\sigma > 0$ is the \textit{scale parameter},
$\theta$ is the \textit{asymmetry parameter} ($|\theta|\leq \min(\alpha,2-\alpha)$) and $\mu \in \mathbb{R}$ the
\textit{shift parameter}.
For example, the Gaussian distribution is found for $\alpha=2$ and has mean equal to $\mu$ and standard deviation equal 
to $\sqrt{2}\sigma$
(the asymmetry parameter $\theta$ is equal to zero when $\alpha=2$).
When $\mu=0$, the class of distributions reduces to the strictly stable distribution $L^\theta_{\alpha,\sigma}$ and
when $\mu=\theta=0$, the class of distributions reduces to the symmetric $\alpha$-stable distributions,
$L_{\alpha,\sigma}$.

\begin{figure}[tb!]
 \centering
\includegraphics[width=0.45\linewidth]{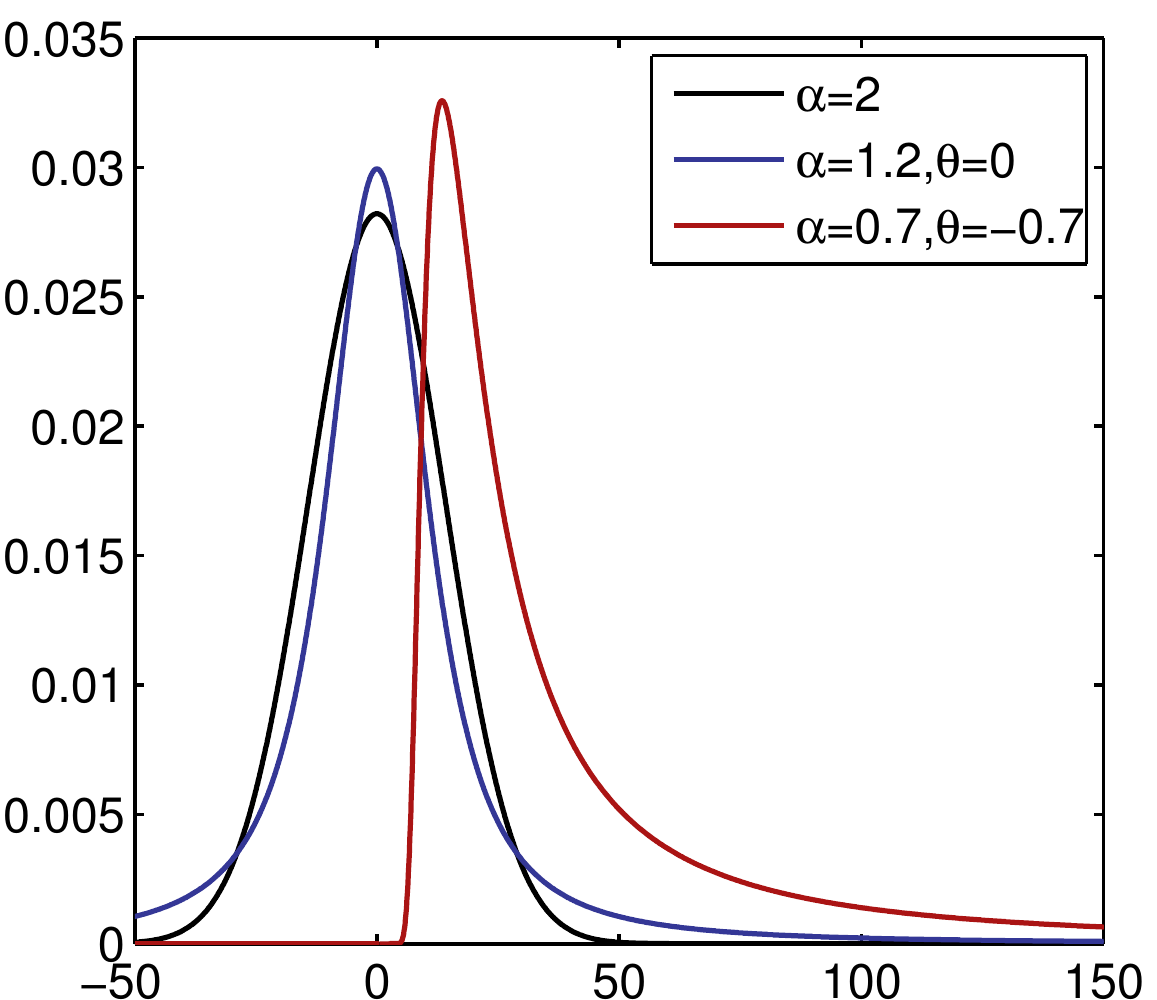}
\includegraphics[width=0.45\linewidth]{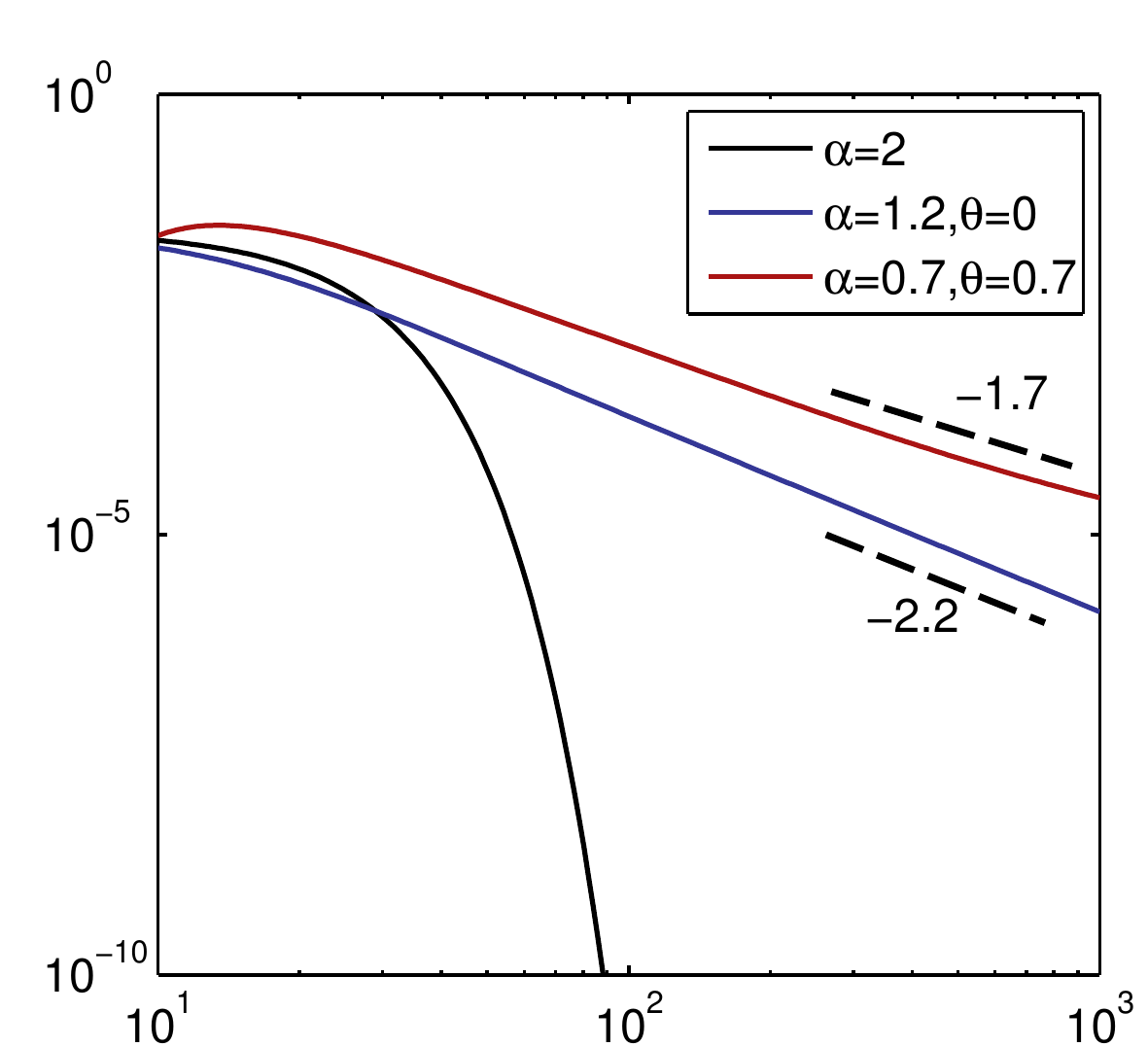}
\caption{Example of stable distributions on a lin-lin (left) and log-log scale (right).
All the distributions have shift parameter $\mu=0$ and scale parameter $\sigma=10$.
A Gaussian distribution is shown in black ($\alpha=2$), a distribution with $\alpha=1.2$ in blue and a one sided
distribution (totally skewed to the right) in red.
The log-log scale shows the heavy tails of the distributions with $\alpha<2$, with exponent $-(\alpha+1)$.}
\label{fig:stable_dist}
\end{figure}

When $\alpha<2$, stable distributions have the interesting property of having one tail, {if their skewness is maximum,}
or both tails that behave asymptotically as power laws (heavy-tail)\cite{Samorodnitsky1994},
\begin{equation}
 L^\theta_{\alpha,\sigma,\mu}(x) \sim \frac{1}{|x|^{\alpha+1}} \mbox{\,, when } \; |x|\rightarrow \infty\mbox{.}
\end{equation}
As a result, they always have infinite variance when $\alpha<2$,
which reflects their capability of modeling processes with large fluctuations.
For $\alpha<1$ they also have infinite first moments.

Stable distributions that have $\alpha<1$ and are totally skewed to the right or to
the left ($\theta = \pm \alpha$) are \textit{one sided}. They are only defined for $x>0$ if $\theta=-\alpha$ and for
$x<0$ if $\theta=\alpha$.

For $\mu=0$ and $\theta=-\alpha$, their Laplace transform is given by\cite{Samorodnitsky1994}
\be
\mathcal{L}[L_{\alpha,\sigma}(x)](s)=\exp\left\{-\frac{\sigma^\alpha}{\cos\frac{\pi\alpha}{2}}s^\alpha \right\}.
\ee
Although no expression in closed form exists for the general stable distributions
{(they are expressed with transcendental functions known as Mittag-Leffler functions\cite{Mainardi2001})},
numerical algorithms
allow to compute their probability distribution function with great accuracy and very efficiently (see \cite{Nolan1997} 
for
example).

\section{Fractional differential operators}
\label{sec:fract_ope}

The idea of generalizing the differential operation to fractional order is as old as differential calculus.
Leibniz, Euler, Liouville, Riemann and Fourier are among the many great mathematicians who developed fractional
differential calculus.
It is only more recently, in the second half of the 20th century, that it began to be applied in physics and
engineering, to problems such as the modeling of viscoelasticity in materials\cite{Caputo1967} or dynamical processes in
fractals\cite{Haber2013}.

Among the different definitions, one of the most famous is the \textit{Riemann-Liouville differential
operators}
that can be defined explicitly by means of the integral operators\cite{Gorenflo2008,Sanchez2005}.
The left and right Riemann-Liouville fraction derivative of order $\alpha$ are
\be
_a\fracdiff_x^\alpha f(x)\equiv\frac{1}{\Gamfnct(m-\alpha)}\frac{\diff^m}{\diff x^m}
\left[\int_a^x\frac{f(x')}{(x-x')^{\alpha-m+1}}\diff x\right]
\ee
and
\be
^b\fracdiff_x^\alpha f(x)\equiv\frac{-1}{\Gamfnct(m-\alpha)}
\frac{\diff^m}{\diff (-x)^m}\left[\int_x^b\frac{f(x')}{(x-x')^{\alpha-m+1}}\diff x\right],
\ee
where $\Gamfnct(t)=\int_0^\infty x^{t-1}e^{-x}\diff x$ is the gamma function, $m$ is the integer satisfying
$m-1<\alpha<m$ and $a$ and $b$ are the start and end point of the operators.
We immediately see from there definitions that an important difference of the fractional version of the derivative  of
a function at a point $x$ is that it is \textit{not a local} property.
As a matter of fact they can depend on the value of the function very far from $x$.
In the cases in which the start point $a$ or the end point $b$ extend all the way to infinity, the following notation
is generally used 
\be
\frac{\diff^\alpha f}{\diff x^\alpha}\equiv_{-\infty}\fracdiff^\alpha_xf(x) \text{\,\,\,\,\,\, and \,\,\,\,\,\,} 
\frac{\diff^\alpha f}{\diff (-x)^\alpha}\equiv^{+\infty}\fracdiff^\alpha_xf(x).
\ee
The Fourier transform of these operators sheds light on their signification and their usage as they appear as a natural
generalization of the Fourier transform of the derivative operator 
\be
\mathcal{F}\left[\frac{\diff^\alpha f}{\diff x^\alpha}\right]=(-ik)^\alpha \hat{f}(k) \text{\,\,\,\,\,\, and
\,\,\,\,\,\,}
\mathcal{F}\left[\frac{\diff^\alpha f}{\diff (-x)^\alpha}\right]=(ik)^\alpha \hat{f}(k).
\ee

A symmetrization of these operators leads to the \textit{Riesz fractional derivative
operator}\cite{Sanchez2005,Calvo2009}
\be
\frac{\diff^\alpha f}{\diff |x|^\alpha}\equiv-\frac{-1}{2 \cos(\pi\alpha/2)}\left(\frac{\diff^\alpha
f}{\diff x^\alpha}+\frac{\diff^\alpha
f}{\diff (-x)^\alpha}\right)
\ee
with the following Fourier transform
\be
\mathcal{F}\left[\frac{\diff^\alpha f}{\diff |x|^\alpha}\right]=|k|^\alpha \hat{f}(k).
\ee

A different definition than the Riemann-Liouville is the \textit{Caputo fractional derivative operator} of order
$\beta$\cite{Caputo1967,Mainardi2001}
\be
_*\fracdiff_t^\beta f(t)\equiv\frac{1}{\Gamma
(m-\beta)}\left[\int_0^t\frac{f^{(m)}(t')}{(t-t')^{\beta-m+1}}dt\right] \text{\,\,\,\,\,\, with \,\,\,\,\,\,}
m-1<\beta<m.
\ee
This definition of the fractional derivative is usually associated with derivatives in time because of the practical
form of its Laplace transform
\be
\mathcal{L}\left[_*\fracdiff_t^\beta f(t)\right]=s^\beta\hat{f}(s)-\sum_{k=0}^{m-1}s^{\beta-k-1}\frac{\diff^k
f}{dt^k}(0)
\ee
which depends only on the initial values of $f(t)$ and its integer derivatives.
The Laplace transform of the Riemann-Liouville derivative depends instead on the initial values of the fractional
derivatives of lower order than $\beta$, which makes is it not practical for real applications\cite{Caputo1967}.

\setlength{\glsdescwidth}{0.9\textwidth}
\printglossary[type=\acronymtype,style=modsuper]

\bibliographystyle{naturemag_nourl}
\bibliography{bibliography} 

\end{document}